\documentclass[12pt]{iopart}
\usepackage{graphicx}
\newcommand{\degs}{\mbox{\(^\circ \)}}

\newcommand{\der}[2]{\frac{d{#1}}{d{#2}}}

\begin{document}

\title[Thermal compensation of test masses]
{Adaptive thermal compensation of test masses 
in advanced LIGO}
\author{Ryan Lawrence, Michael Zucker, Peter Fritschel, Phil Marfuta 
and David Shoemaker}
\address{Department of Physics and Center for Space Research, 
Massachusetts Institute of Technology, Cambridge, MA  02139}
\vspace*{5pt}\address{E-mail: {{\tt ryan@ligo.mit.edu}}}

\begin{abstract}
As the first generation of laser interferometric gravitational wave detectors
near operation, research and development has begun on increasing the
instrument's sensitivity while utilizing the existing infrastructure.  
In the Laser Interferometer Gravitational Wave Observatory (LIGO), 
significant improvements are being planned for installation in 
$\sim$2007, increasing strain sensitivity through improved suspensions and 
test mass substrates, active seismic isolation, and higher input laser power.  
Even with the highest
quality optics available today, however, finite absorption of laser power
within transmissive optics, coupled with the tremendous amount of optical power
circulating in various parts of the interferometer, result in critical 
wavefront deformations which would cripple the performance of the instrument.
Discussed is a method of active wavefront correction via direct thermal
actuation on optical elements of the interferometer.    A simple nichrome
heating element suspended off the face of an affected optic will, through
radiative heating, remove the gross axisymmetric part of the original thermal
distortion.  A scanning heating laser will then be used to remove any remaining
non-axisymmetric wavefront distortion, generated by inhomogeneities in the
substrate's absorption, thermal conductivity, etc.  A proof-of-principle
experiment has been constructed at MIT, selected data of which are
presented.
\end{abstract}

\submitto{\CQG}
\pacs{04.80.Nn, 95.55.Ym, 42.60.Lh, 07.60.Ly}

\maketitle

\section{Introduction}

As the first generation of the Laser Interferometer Gravity Wave Observatory 
(LIGO)\footnote{http://www.ligo.caltech.edu} nears operation, planning is 
underway to further improve the 
instrument's phase sensitivity.  Around 2007, new active seismic isolation
systems, new test masses and suspensions, and more powerful lasers will all be 
installed in the existing infrastructure, with the goal of 
increasing the broadband phase sensitivity (hence gravitational wave strain 
sensitivity) of the instrument by at least a factor of 10.  The primary 
contributor to phase 
noise at gravitational-wave frequencies above 300 Hz is the 
Poisson fluctuation in detected laser power, termed 
``shot noise''.  
In principle, one can increase the optical power in the interferometer to 
reduce this effect, but only to the fundamental limit where 
fluctuating radiation pressure 
randomly perturbing the test mass momenta (a noise source whose amplitude 
increases with power) overtakes the shot noise, an effect enforced by the 
Heisenberg Uncertainty limit for detection of the mirror position.

\begin{figure}[t]
\centerline{\scalebox{.5}{\includegraphics{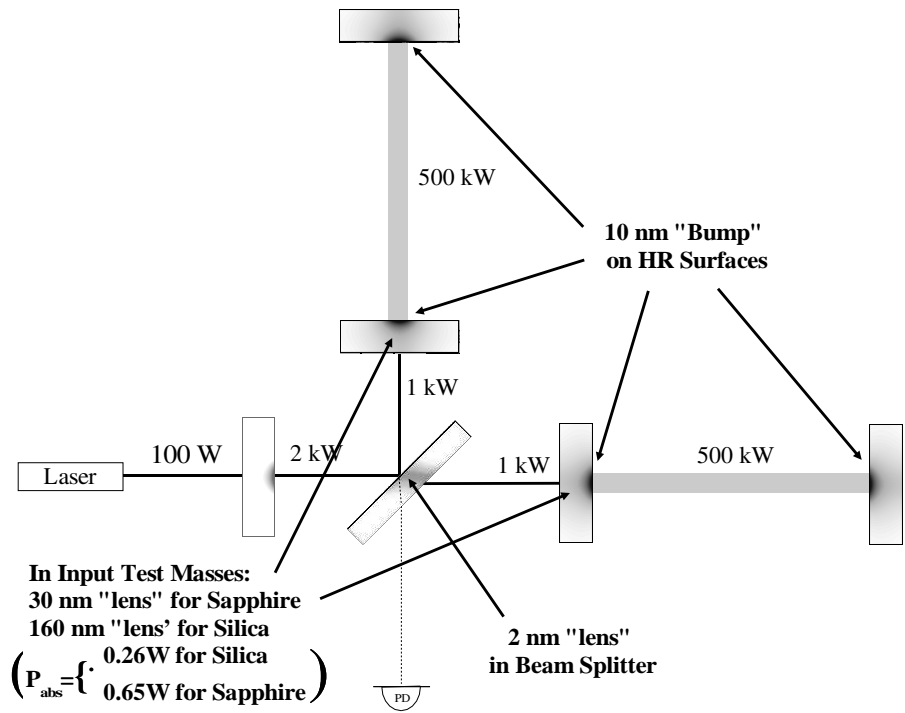}}}
\caption{\label{F:Overview} Projected thermal effects in advanced LIGO.}
\end{figure}

However, the amount of allowable circulating optical power is
limited by
the nonzero optical absorption in the substrate and coatings of test 
masses, typically 0.5 parts per million (ppm) per reflection for coatings 
and 1 ppm/cm for transmission through
fused silica substrates.  Absorption of the 
Gaussian-profiled beam induces a 
nonuniform temperature increase within the optic.  This induces 
nonuniform optical path length distortions through two mechanisms: 
(1) thermoelastic 
expansion of the optic's surface, and (2) variation of the material's 
refractive index with temperature (termed ``thermal lensing'').  One may
approximate that, for 
a collimated Gaussian beam propagating through a cylindrical optic, the 
absorption-induced optical path difference between the beam center and 
$1/e^{-2}$ intensity radius (termed the ``waist radius'' $w$) is 
\cite{Germans}:
\begin{equation}\label{E:approx}
\eqalign{
\delta &s \approx P_{abs} \frac{ \beta }{ 4 \pi \kappa } \approx
6\times10^{-7}\frac{\mbox{m}}{\mbox{W}}\,
\left( \frac{1.38\,\frac{\mbox{W}}{\mbox{m}\,\degs{\mbox{K}}}}{\kappa} \right)
\left( \frac{\beta}{1\times10^{-5} /\degs{\mbox{K}}} \right) \, P_{abs}\\
\mbox{\hspace{-2cm} where} \quad &\beta=
\cases{\alpha+\der{n}{T},& in transmission \\
       \alpha,           & on reflection   \\}}
\end{equation}
where $\kappa$ is the thermal conductivity, $\alpha$ is the 
thermal expansion coefficient, and $\der{n}{T}$ 
is the derivative of the index of refraction with respect to temperature 
for the substrate material, and $P_{abs}$ is the optical power absorbed 
in the optic (in both the substrate and the coatings).  Figure 
\ref{F:Overview} details the expected distortions in an advanced LIGO 
interferometer.

To enable readout and control of the LIGO interferometer's mirror 
positions, a deliberate arm length asymmetry of about 22 cm is introduced 
to the 
power recycling cavity (a nearly degenerate cavity consisting of the 
recycling mirror and the high reflectivity surface of the two arm cavity 
input couplers), and radio-frequency (RF) sidebands are impressed on the input 
laser light which resonate in the Michelson cavity and are anti-resonant in 
the arm cavities \cite{Gang1}.  With
this in mind, the deleterious effects of thermal lensing and 
thermoelastic expansion are seen to be twofold.  Firstly, thermal lensing in 
the beamsplitter, as well as distortions common to both arms (due to the 
deliberate arm length asymmetry), result 
in imperfect interference at the antisymmetric port; this results in 
additional light on the photodetector which 
contains no information 
about the instrument's differential arm length, but contributes shot noise 
nonetheless.  Secondly, thermal lensing in the input test masses can cause the 
power recycling cavity to become unstable for the RF 
sidebands.

To calculate the effects thermal distortions have on the 
circulating optical fields in the instrument we use
the Melody\footnote{available at the URL 
http://www.phys.ufl.edu/LIGO/LIGO/STAIC/SOFT/}
 modal model of the LIGO interferometer \cite{Ray,Ray1}.  
Knowing the fields in the modeled interferometer, we calculate the 
shot noise limited phase sensitivity as \cite{Gang}:
\begin{equation}\label{E:shotnoise}
\Delta \tilde{\phi}(f)=
\left(F_{ns}\sqrt{ 1+ R_{bs}\frac{1-C}{4\sin^2(2 k_m\delta l)}} \right)\,
\sqrt{\frac{2 \hbar \omega}{P_{bs}}}
\end{equation}
where $F_{ns}$ is the non-stationary correction factor ($\sqrt{3/2}$ for
the current readout scheme), $k_m=2\pi f_m/c$ 
is the modulation wavenumber for the resonant sidebands, $R_{bs}$ is the 
ratio of carrier power to sideband power at the beamsplitter, 
$C\equiv(P_{bs}-P_{as})/(P_{bs}+P_{as})$ is the fringe contrast, $\omega$
is the laser light frequency, $P_{bs}$ is the carrier power incident on 
the beamsplitter, and $P_{as}$ is the total carrier power at the 
antisymmetric port.  Figure
\ref{F:Mldyshot} shows the carrier power expected at the antisymmetric 
port, as well as the resulting phase noise, as a function of interferometer 
input power.

\subsection{Thermal compensation}

\begin{figure}[t]
\centerline{\scalebox{.5}
{\includegraphics[angle=-90]{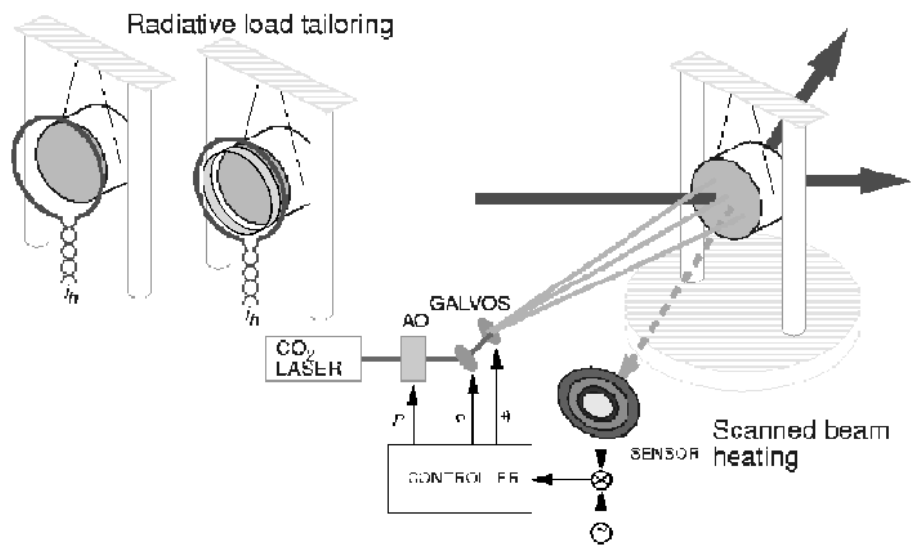}}}
\caption{\label{F:cartoon} Compensation methods.}
\end{figure}

Research is underway on techniques of avoiding the problem of power
absorption in LIGO optics.  Identifying and reducing the source of 
bulk optical absorption in test mass material is one obvious solution 
\cite{Abs}.  One may also change the topology of the interferometer
to an all-reflective one \cite{Beyersdorf}, which exploits the fact that 
thermal expansion coefficient for fused silica is $\sim 20 \times$ smaller 
than its $\der{n}{T}$.  We discuss a method of directly addressing the 
problem by ``thermal compensation'':  one may attempt to
homogenize the absorption-induced temperature fields by radiatively 
depositing additional heat on the surface of the optic in a tailored 
pattern, thus compensating the original distortion.  We 
investigate
two methods for thermally compensating distortions in the LIGO interferometer.
First, one may arrange radiative heating elements and reflectors to thermally 
compensate 
the anticipated beam-induced thermal distortions.  This method is expected to 
work well for distortions whose form can be calculated in advance.  More 
generally, a scanning laser beam with a wavelength which is strongly absorbed 
by the optic's substrate can be used to 
actively tailor the heating pattern to address non-axisymmetric distortions 
induced by possible optical inhomogeneities and surface impurities.  Figure
\ref{F:cartoon} sketches the two modes of thermal compensation we investigate.

To determine the effects of a given heating pattern on the surface of
the optic, we have constructed a two dimensional finite element model 
which, given a one dimensional heating pattern (axisymmetric heating), 
numerically solves the partial differential equations (PDE's) governing 
heat transfer, then uses the calculated temperature field to solve the 
PDE's governing thermoelastic deformations.  As a figure of merit for 
a given correction, we use the resulting corrected wavefront distortion to 
calculate the total power scattered out of the incident TEM$_{00}$ LIGO 
beam into higher order spatial modes upon reflection or transmission by
the corrected optic.  A second practical figure of merit is the maximal 
temperature increase of the optic, including the heat added by our thermal
correction, per Watt of absorbed optical power compensated.

Finally, we note that it is not entirely 
necessary to directly thermally actuate the optic whose distortions we
wish to correct, but may instead impose the correction on a separate, purely
transmissive compensation plate placed in the optical chain.  Sapphire, the 
material proposed for the advanced LIGO test masses, has a thermal conductivity
$\sim 30\times$ larger than fused silica as well as a high transmission 
for wavelengths shorter than $6\,\mu$m:  two properties which 
make sapphire undesirable compared to fused silica for direct thermal 
actuation.  For the remainder of this paper, we discuss the effects
of thermal compensation on fused silica optics only.

\section{Axisymmetric thermal compensation}

We first consider a thin heating ring, mounted 
at a distance $H$ from the face of the optic and axially centered on 
the optical axis (symmetry axis) with a radius $R$ larger than the 
optic's radius (to 
prevent interference in the clear aperture)).  Inserting the calculated 
heating pattern into our 
finite element model to calculate wavefront corrections, and optimizing over 
$R$, $H$, and the ring power $P_{ring}$, we find a 100-fold reduction in the 
power scattered into higher order modes.  However, this comes at the cost of 
a temperature increase of $375\degs$K per Watt of optical power compensated.  
The simple ring is thus deemed impractical for realistic thermal compensation.

One clear contributor to the inefficiency of the single heating ring is the
fact that some of its radiation falls in the center of the optic, where the 
beam heating
which we are attempting to compensate is concentrated.  One method of 
eliminating this spurious central heating is to include a 
passive cylindrical shield, with a radius 
equal to the optic's (thus smaller than the ring's), and a height $H_s<H$
such that the central portion of the optic is shielded from the ring's 
radiation.  Another contributor to the inefficiency of the single heating 
ring is the dissipation of heat out the radial surface of the optic.  
Suspending a low-emissivity aluminum sheath around this surface inhibits 
radiation exchange with the environment, hence reducing radial temperature 
gradients.

\begin{figure}[t]
\centerline{
\scalebox{.5}{\includegraphics{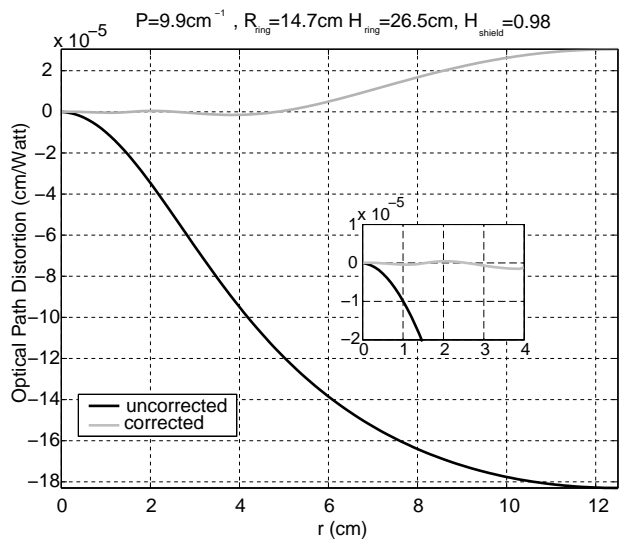}}}
\caption{\label{F:shieldring} Optical path distortion through a fused 
silica input test mass, per Watt of absorbed optical power.}
\end{figure}

Again using our finite element model to optimizing over 
$R$, $H$, $H_s$, and $P_{ring}$, we find a 1000-fold reduction in TEM$_{00}$ 
scatter is achievable over a broad range of $R$ and $H$, at the cost of a
temperature increase in the substrate of $48\degs$K per watt of optical 
power compensated.  
Figure \ref{F:shieldring} shows an optimized corrected wavefront 
for the shielded ring actuating an insulated fused silica input test mass of 
LIGO I dimensions (radius 12.5 cm, height 10 cm, beam waist 3.6 cm).

\begin{figure}[t]
\centerline{\scalebox{.5}{\includegraphics[angle=90]{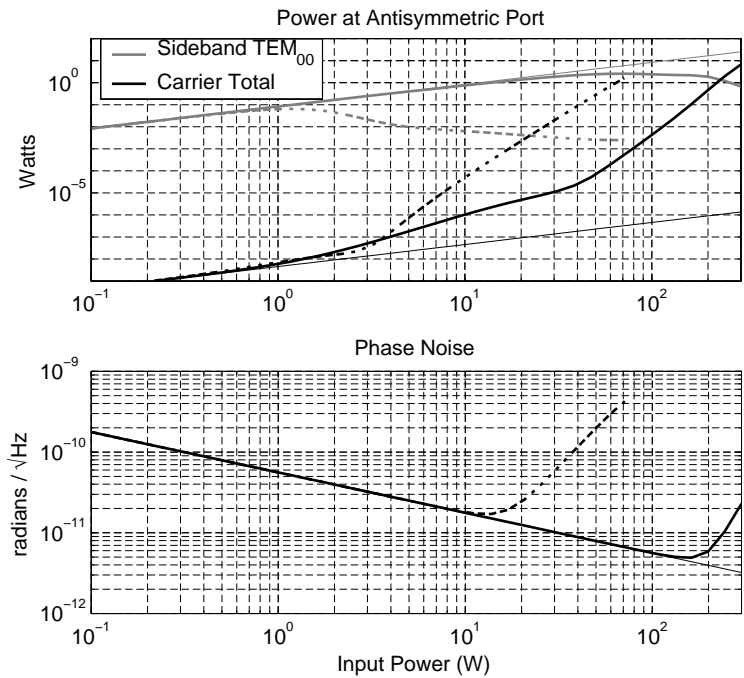}}}
\caption{\label{F:Mldyshot} Modeled level of output carrier and sideband 
powers, as well as the detected phase noise 
(shot noise component) versus optical power at the interferometer input.  The 
broken line is the uncompensated case, the bold line is compensated, and the 
narrow lines are without any thermal distortions.}
\end{figure}

Inserting the corrected wavefront distortions for an input test mass into 
the Melody model of the interferometer, we calculate the effects of 
thermally actuating the test masses on the circulating fields of the 
interferometer.  Calculating the phase noise component, we further optimize
to compensate the beamsplitter lens by adjusting the correction 
amplitude (i.e., the heating ring power) on the test masses.  
Figure \ref{F:Mldyshot} shows 
the instrument's performance with the input test masses compensated
as detailed in figure \ref{F:shieldring}, with a relative compensation 
amplitude of 1.00 in the X arm input test mass, 0.95 in the Y arm input 
test mass.

\section{Scanning laser thermal compensation}

In the case where the material properties of our optics (absorption, thermal
conductivity, etc.) are not perfectly homogeneous, additional non-axisymmetric 
distortions may need to be compensated.  We use a well absorbed laser beam, 
amplitude modulated and scanned over a single face of an actuated optic.  By 
tailoring the power 
deposited in over a discrete, constant scanning pattern with a scan period 
small compared to the local thermal time constant $\tau_l$ ($\sim
w_h^2 \rho c/\kappa$ where $\rho$ the density, $c$ the heat 
capacity, $\kappa$ the thermal conductivity of the optic, and $w_h$ the heating
beam waist radius), one can compensate a subset of all possible wavefront 
distortions. 

Given a fixed scan pattern of $N$ points, we wish to find the power which
must be deposited at each scan point, represented as a vector $\vec{P}$ of 
length $N$, to produce a given wavefront distortion $\vec{d}$, 
represented as a vector in some $M$ dimensional basis spanning a subset of 
$\mathbf{L}^2(D)$, the space of integrable 2D functions over a region $D$ 
(defined as a circular aperture of radius $R$).  We can calculate (or 
measure) the $N\times M$ ``actuation matrix'' $\underline{A}$ defined such 
that:
$$\underline{A}\cdot\vec{P}=\vec{d}.$$ 
Inverting $\underline{A}$ (or using the technique of Single Value Decomposition
in the case that $\underline{A}$ is not square \cite{Tyson}), we get 
the $M\times N$ ``response matrix'' $\underline{R}=\underline{A}^{-1}$ such 
that:
$$\underline{R}\cdot\vec{d}=\vec{P}.$$
Thus, given a wavefront distortion $\vec{d}$, we arrive at the power
vector required to produce it.

Instead of working in a basis of orthogonal polynomials, it is 
perhaps simpler to work in the basis of ``actuation functions'' 
themselves (the $k$th actuation function $A_k(r,\theta)$, 
is the net distortion generated per unit power by the laser actuating on the 
$k$th scan point), which are, in general, not orthogonal in the normal 
$\mathbf{L}^2$ sense.  As long as each function $A_k(r,\theta)$ is 
linearly independent 
of the previous $k-1$ functions, it is possible to construct an orthogonal 
basis from the $A_k$ that spans the space of compensatible wavefronts for a 
given scan pattern (see, for example, \cite{math}).  Hence, we may decompose 
a given wavefront $\Phi(r,\theta)$ directly into our non-orthogonal basis 
$\{A_k\}$ with the standard normalized inner product, i.e.,:
$$d_k=\frac{\int \int A_k(r,\theta)\Phi(r,\theta)\,r\,dr\,d\theta}
{\int \int A_k^2(r,\theta)\,r\,dr\,d\theta}$$
and clearly $P_k=d_k$.

This result is, in general, not positive for all $k$, which is 
problematic since we cannot actuate on an optic with negative power.  
The direct way to bypass this is to utilize the fact that a 
``deformation'' which is constant over the aperture $D$ is compensatible by 
simply moving the optic we wish to compensate.  Adding
an arbitrary constant $C$ such that $\Phi+C\ge0$ over the
entire aperture ensures that all the $P_k$ are $\ge 0$.  However, to 
fully span the subspace of compensatible wavefronts, we must be able to
generate $-A_k$ just as well as we can generate $A_k$.  More generally, we 
must be able to generate all linear combinations \cite{math}:
$$c_1 A_1+c_2 A_2+\dots+c_N A_N.$$
Again utilizing the fact that we do not need to compensate ``distortions'' 
which are constant over the aperture $D$, an additional (sufficient) condition 
for our basis of actuation functions to fully span the subspace is that:
$$C=\sum_{n=1}^N A_n(r,\theta)$$
where $C$ is some constant.  The fulfillment of this condition puts 
a rigorous constraint on the scan patterns we may choose.

\section{Thermal compensation experiment at MIT}

\begin{figure}[t]
\centerline{\scalebox{.5}{\includegraphics[angle=90]{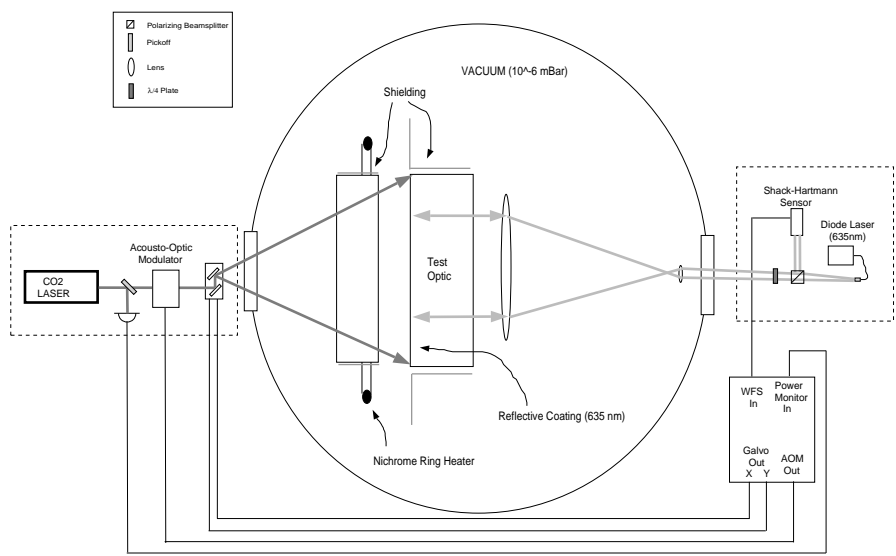}}}
\caption{\label{F:schematic} Schematic of the experiment.}
\end{figure}

Figure \ref{F:schematic} diagrams the experiment at MIT to experimentally 
test the 
basic principles of both axisymmetric and scanning-laser thermal compensation.
 We use a 
Shack-Hartmann wavefront sensor to detect changes in optical path when we 
thermally actuate the test optic through either a scanned CO$_2$ laser or a 
shielded heating ring.
The optic under test is radially insulated with aluminum foil, and mounted in 
high vacuum ($1\times10^{-6}$ mBar). The probe source is a fiber pigtailed 
diode laser 
($\lambda=635$nm), coupled into a single mode fiber and collimated with a 
grin lens.  The beam is expanded to the size of the optic under test, 
which is mounted in a high vacuum chamber.  The reflected 
beam from transmission through the optic is sent back through the 
optical system, where is it diverted with a quarter-waveplate/polarizing
beamsplitter and directed into a commercial Shack-Hartmann 
wavefront sensor.  Wavefront slopes relative to the initial 
``cold'' wavefront are resolved over the Shack-Hartmann sensor's 
$32\times24$ lenslet grid, and the resulting optical path distortion is 
reconstructed over the clear aperture.

\begin{figure}[t]
\centerline{\scalebox{.5}{\includegraphics{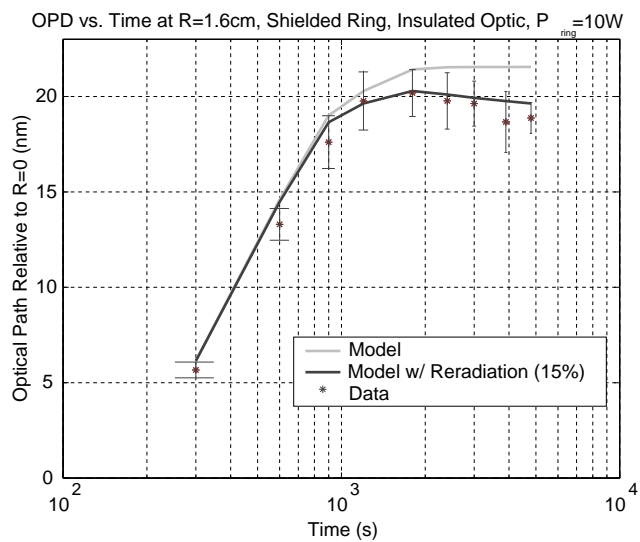}}}
\caption{\label{F:ringdata} Optical path distortion at $r=16$mm versus 
time for a shielded ring actuator on a fused silica test optic ($R=5$cm, 
$H=8$cm).}
\end{figure}

Initial data taken for the shielded ring actuator shining on 
a fused silica test optic (5 cm radius, 8 cm depth) can be seen in 
figure \ref{F:ringdata}, and are in good agreement with 
theory.  However, a decay in 
the correction magnitude was seen over time, mainly due to re-radiation 
of the heat shield which was not thermally grounded in an adequate manner.

\begin{figure}[t]
\centerline{\scalebox{.5}{\includegraphics{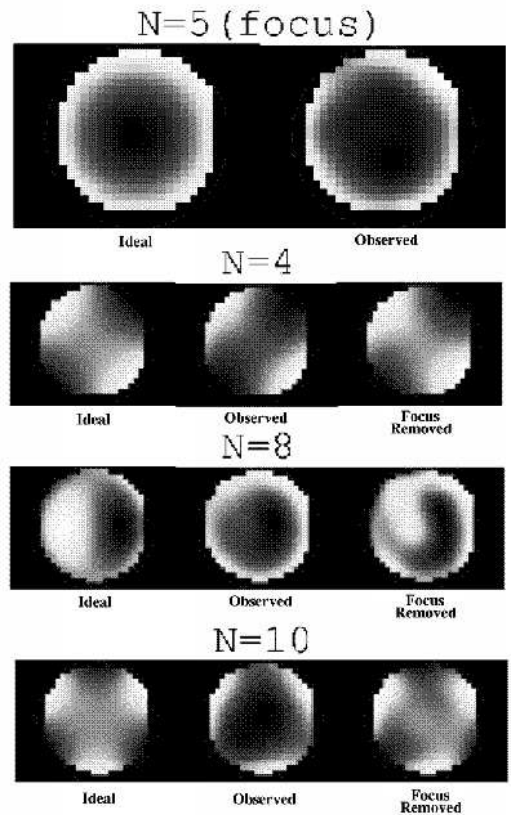}}}
\caption{\label{F:prettyzern} Zernike polynomials experimentally replicated 
by transmission through a fused silica test optic.}
\end{figure}

Initial tests have been completed on the same fused silica test optic for 
the scanning laser \cite{Phil}, and select data can be seen in figure 
\ref{F:prettyzern}.
Here we have used the basis of Zernike polynomials to numerically compute 
the response matrix, as we have discuss in the first part of section 3.  The 
aperture is 2.5 cm in 
radius, and the actuating beam has a waist radius 0.5 cm.  Our initial 
calculations disregarded thermal expansion terms in the actuation matrix, 
as $\der{n}{T}$ is about $10\times$ larger than $\alpha$ in fused silica.  
 The spatial frequency of the recreated wavefronts was limited by the 
beam size in this test, and we could not reproduce Zernike terms higher than 
$N=10$ (i.e., $Z_{3\,3}$).  Also, the net positive amount of power deposited 
on a single face of the optic caused it to thermoelastically bow, thus giving 
a persistent curvature term in these data.
Since we are working in the regime of small temperature increases, as well
as small physical distortions, taking thermal expansion terms into account 
when calculating the actuation functions will eliminate the persistent 
curvature.
 
\section{Conclusions}

We have discussed the numerical and analytical work done toward solving the 
problem of thermally compensating the thermal distortions induced in the 
next generation of LIGO.  Using our numerical model to find an optimum 
correction for a given distortion, we insert the results into a modal
model of the entire interferometer and find that a shielded ring heater
actuating on the recycling cavity side of the arm cavity input couplers can
facilitate the proposed operating power of advanced LIGO with fused silica
test masses.  In addition, we have begun both analytical and experimental
work on developing a scanning laser actuator to compensate any non-axisymmetric
wavefront distortions that may be present themselves in the actual 
advanced LIGO optics.

\ack

We thank R. Weiss and P. Schechter for useful discussions and ideas, and are 
especially grateful to R. Beausoleil for the use of and technical advice 
regarding Melody.  This work was supported by NSF grant PHY-9210038.

\end{document}